\documentstyle[12pt]{article}
\def\gev{{\rm \, Ge\kern-0.125em V}}
\begin{document}
\begin{titlepage}
\pagestyle{empty}
\baselineskip=21pt
\rightline{McGill 95--46}
\rightline{UMN--TH--1414/95}
\rightline{hep-th/yymmddd}
\rightline{October 95}
\vskip .2in
\begin{center}
{\large{\bf Axions and the Graceful Exit
Problem in String Cosmology}}
\end{center}
\vskip .1in
\begin{center}
Nemanja Kaloper

{\it Department of Physics, McGill University}

{\it Montr\'eal, Qu\'ebec, Canada H3A 2T8}

and

Richard Madden and Keith A. Olive

{\it School of Physics and Astronomy, University of Minnesota}

{\it Minneapolis, MN 55455, USA}

\vskip .1in

\end{center}
\vskip .5in
\centerline{ {\bf Abstract} }
\baselineskip=18pt
We reexamine the graceful exit problem in the
 Pre-Big-Bang inflationary scenario.
The dilaton-gravity action is generalized by
adding the axion and a general
axion/dilaton potential. We provide  a
phase space analysis of the dynamics which leads us to
extend the previous no-go theorem and
rule out the branch change necessary for
graceful exit in this context.

\vskip.5in
\centerline{\it Submitted to Physics Letters {\bf B}}

\end{titlepage}
\baselineskip=18pt
{\newcommand{\la}{\mbox{\raisebox{-.6ex}{~$\stackrel{<}{\sim}$~}}}}
{\newcommand{\ga}{\mbox{\raisebox{-.6ex}{~$\stackrel{>}{\sim}$~}}}}
\def\beq{\begin{equation}}
\def\eeq{\end{equation}}
\section{Introduction}

The standard cosmological model provides a convincing and
consistent picture of the universe back to the period of primordial
nucleosynthesis. When extrapolated further into the past, however,
it reveals a need for a very special set of initial conditions,
both finely tuned (the flatness problem) and coherent over acausal
scales (the horizon problem). Together with other well known
problems and gaps, not the least of which is that the model predicts
its own demise by necessitating a past
singularity \cite{sing}, these problems justify the
search for mechanisms in the early universe which
can erase deviations from special initial conditions, providing
a dynamical solution to the fine tuning. The inflationary
paradigm \cite{infl} does just this despite the fact that there
is no consensus on a concrete
mechanism.

Since string theory \cite{string} is the only current candidate for a theory
capable of uniting gravity with the other forces of nature, there
have been many investigations of the effects of stringy dynamics
on cosmological evolution \cite{dilinf}-\cite{mv}. In particular,
the massless modes (dilaton, axion and graviton) of
the string can develop classical fields whose couplings are fixed
by the requirement of conformal invariance, leading to the
low energy effective action for these fields.

In addition to providing
new evolutionary possibilities, solutions to these equations have
a larger symmetry group \cite{bvafa}-\cite{mv}
including scale-factor duality ($r(t)
\rightarrow 1/r(t)$, where $r(t)$ is the scale factor). Combining
this with time reversal symmetry, one can find that corresponding
to an era of decelerated expansion defined for positive time there
is another solution representing accelerated expansion (inflation)
defined for negative times. Assuming these could
be connected, one would obtain a cosmological solution
describing a slowly expanding universe which accelerates
into a period of rapid expansion, identified as the
`big bang', and then settles into a standard Friedmann-Robertson-Walker
cosmology. This is the `pre-big-bang' scenario \cite{gv,bv}.

To promote this idea into a realistic
model we are still lacking two
essential components. First, the stringy fields must be tamed
in the post-big bang universe. In particular, the dilaton
must be decoupled since variations in the dilaton field
correspond to changes in masses and coupling constants,
which are strongly constrained by
observation \cite{dp,co}. This
can be accomplished by including dilaton self-interaction
potentials and trapping the dilaton in a potential minimum.
While the introduction of a general potential will destroy
exact scale factor duality symmetry we can expect the
asymptotic forms of the solutions to remain similar
in regions where the potential tends to a constant value.
Second, we need to
check if the symmetry related branches can be linked.
Previously we followed a suggestion \cite{bv} that the
potential itself could catalyze this branch changing,
though in \cite{bv} the authors in fact noted the difficulty
of achieving such a branch change and claimed a hard-to-go
theorem obstructing the change.
In \cite{lastkmo}, we confirmed their negative conclusions
by proving an exact no-go theorem for the correct form of
branch changing by a dilaton potential,
even with a stringy fluid or higher genus terms as
suggested in \cite{dp}.

Recently there have been some speculations about the fate of
the no-go theorem when axions, with or without
self interactions, are present. Namely, it is known that the
axion terms could affect dynamics of collapsing universes,
and actually overturn collapse into expansion \cite{clw},
eventually asymptotically linking separate solutions
of the axionless theory. Thus there may have been hope
that, with the inclusion of the axion interactions, a similar
link-up between expanding solutions belonging to different
branches might occur. This turns out not to be the case.
In this letter we extend the dilatonic no-go result by including
the axion and an arbitrary dilaton/axion potential and showing
that one cannot connect solutions well-behaved in the past
with those well-behaved in the future.
The proof is a direct analog of the proof for the case of a
potential depending only on the dilaton.

\section{The Gravitational Action}

There is a considerable amount of literature
concerning the tree level
gravitational action in string theory and its
expansion in the string tension
$\alpha'$ \cite{acts}.
In what follows we will ignore the corrections of the
derivative expansion of order $O(\alpha')$ and higher.
Though these terms would be
expected to be important near the singularity,
we will neglect them for the
reasons of simplicity and consistency (these
terms would involve higher
derivative corrections in the equations of motion,
including spurious
solutions which must be non-perturbative
in $\alpha'$ by
dimensional analysis \cite{simon}).
Retaining the conventions of \cite{lastkmo},
in the string world-sheet frame the tree-level action
becomes:
\beq
S = \int d^4x \sqrt{g} e^{-2 \phi} \Bigl\{ R
 + 4 (\nabla \phi)^2 - \frac{1}{12} H_{\mu \nu \lambda}
H^{\mu \nu \lambda}
  \Bigr\}
\label{sact1}
\eeq
where the 3-form $H=dB$ is the field strength of the
Kalb-Ramond 2-form
$B_{\mu \nu}$. In four dimensions, this 3-form is
dynamically dual to a
pseudoscalar axion field. The correspondence
is given through the Hodge
duality to a vector $Y$
\beq
H_{\mu \nu \lambda}=\sqrt{g}
   \epsilon_{\mu \nu \lambda \rho} Y^{\rho}
\eeq
which has to be constrained further in order to satisfy the
Bianchi identity $dH=0$. This is ensured by adding
a Lagrange multiplier term of
the form $a(dH)$ to the action, and
integrating out the 3-form $H$ and the 1-form $Y$, in favor of
the pseudoscalar degree of freedom $a$. The final
correspondence is given by
\beq
H_{\mu \nu \lambda}=\sqrt{2} e^{2 \phi} \sqrt{g}
   \epsilon_{\mu \nu \lambda \rho} \partial^{\rho} a
\eeq
resulting in the replacement of the 3-form kinetic term
coupled to
the inverse string coupling $e^{-2 \phi}$ by the pseudoscalar kinetic term
coupled to the dual string coupling.

Finally, motivated by the idea that a potential
could simulate generic properties of branch changing,
we can introduce a potential for the dilaton and
axion fields, coming from the loop expansion
of the effective action
(we will not speculate as to the origin of
this potential and leave it
completely general). This yields the following action:
\beq
S = \int d^4x \sqrt{g} e^{-2 \phi} \Bigl\{ R
 + 4 (\nabla \phi)^2 - e^{4 \phi} (\nabla a)^2 -
2 \Lambda(\phi,a)  \Bigr\}
\label{sact2}
\eeq

\def\dldp{{{\partial \Lambda} \over {\partial \phi}}}
\def\dlda{{{\partial \Lambda} \over {\partial a}}}
\def\dldpda{{{\partial^2 \Lambda} \over {\partial a \partial \phi}}}

We shall be working with the FRW metric in the
string frame,
\beq
ds^2=-n(t)^2 dt^2+r(t)^2 d{\vec x}^2
\eeq
where $d{\vec x}^2$ is the three dimensional volume
element for a space of
constant curvature $k$. We will concentrate on the case
of $k=0$. To be consistent with this form of the
metric we will also assume that the dilaton and axion
field are functions only of time. To obtain the equations
of motion we can work directly with the action.
Expressing the action in terms of
the functions $r(t)$ and $n(t)$ we can
Kaluza-Klein reduce it to a one-dimensional problem in
constrained mechanics,
and look at the variations of the resulting action
with respect
to the functions $n(t)$, $a(t)$, $\phi(t)$ and
$r(t)$, finally
setting $n(t)=1$. The equations of motion which
arise from these
variations are respectively
\begin{eqnarray}
\Lambda + 6 \dot \phi h - 3 h^2 - 2 {\dot \phi}^2
  + {1 \over 2} e^{4 \phi} {\dot a}^2 &=& 0 \label {e1}
\\
\noalign{\medskip}
{\ddot a} + 3 h \dot a+ 2 \dot \phi \dot a + e^{-4 \phi}
   \dlda &=& 0 \label {e2}
\\
\noalign{\medskip}
4(\ddot \phi-{\dot \phi}^2 + 3 h \dot \phi) -
6 (\dot h + 2 h^2) +  2 \Lambda - \dldp +
   e^{4 \phi} {\dot a}^2 &=& 0 \label{e3}
\\
\noalign{\medskip}
4 (\ddot \phi-{\dot \phi}^2 + 2 h \dot \phi) -
4 \dot h - 6 h^2 - e^{4 \phi} {\dot a}^2 +
2 \Lambda &=& 0 \label{e4}
\end{eqnarray}
where $h=\dot r/r$.
Multiplying the axion equation of motion (\ref{e2})
by $\dot a$
we can recast it in terms of the variable
$\rho=e^{4 \phi} {\dot a}^2$. By further noting that
\beq
\dlda \dot a = {\dot \Lambda} - \dldp \dot \phi
\eeq
we can rewrite this equation as:
\beq
\dot \rho + 6 h \rho + 2 \dot \Lambda -
  2 \dldp \dot \phi = 0
\eeq
Manipulating (\ref{e3}) and (\ref{e4}) to eliminate
$\ddot \phi$ and ${\dot \phi}^2$ and retaining
the constraint equation ($\ref{e1}$) we finally obtain
\begin{eqnarray}
\dot h &=& 2 h \dot \phi - 3 h^2 + \rho -
   (\dldp)/2 \label{ee1}
\\
\noalign{\medskip}
0 &=& \dot \rho + 6 h \rho + 2 \dot \Lambda -
  2 \dldp \dot \phi \label{ee2}
\\
\noalign{\medskip}
0 &=& 2 {\dot \phi}^2 + 3 h^2 - 6 h \dot \phi -
\Lambda - \rho / 2
\label{ee3}
\end{eqnarray}
which is the set of equations of motion we wish to analyze.

\section{Branch Changing and the Graceful Exit}

To investigate the dynamics of the trajectories we
consider motion
in the four dimensional phase space of the
variables $h, \phi, a$ and
$\rho$.
The constraint equation (\ref{ee3}) can
be solved for $\dot \phi$
\beq
\dot \phi={{3 h \pm \sqrt{3 h^2 + 2 \Lambda + \rho}} \over 2}
\label{dotphi}
\eeq
To be consistent with previous work \cite{bv,lastkmo}
we should designate the trajectories
having the plus/minus sign in the above quadratic as (+)/($-$)
branch trajectories. We will also need the
egg function
\beq
e=\sqrt{3 h^2 + 2 \Lambda + \rho}
\label{eggdef}
\eeq
where we note that trajectories can switch branches only
when $e=0$, because only here $\dot \phi$ is
identical on both branches.
(In fact, branch changing must generally occur here to keep
higher derivatives continuous). The region where $e \le 0$ is
referred to as the ``egg''. Clearly an egg can
occur only where $\Lambda \le 0$.

By substituting (\ref{dotphi})
into (\ref{ee1}) we get the evolution
equation for $h$:
\beq
\dot h = \pm h \sqrt{3 h^2 + 2 \Lambda + \rho} + \rho -
   (\dldp)/2
\label{doth}
\eeq
We can now see that the
(+) branch solutions are susceptible
to runaway expansion when $h$ dominates
this equation whereas
the ($-$) branches are stable against
this type of future behavior.
This in turn indicates that a (+) will tend
to evolve out of
a well-behaved past into a future
singularity and the reverse
for a ($-$) branch. Thus a (+)
into ($-$) transition, if
dynamically allowed, could
be a globally nonsingular cosmology
as indicated in the introduction,
having a region of very large
curvature mimicking the Big Bang,
and possibly an inflationary phase, in the
neighborhood of this transition.

Now, as we have shown before, in the pure
dilaton-metric system
such transitions are impossible.
However, referring to (\ref{ee2}),
we see that in the simple case of an
axion-independent potential,
$\rho \propto r^{-6}$. This in fact means
that a non-zero axion can prevent the universe from
a singular collapse, as shown in \cite{clw}.
However, the free axion
(note the tentative use of the label
``free" here, meaning axion-independent potential;
the axion still
couples exponentially to the dilaton)
redshifts away rapidly in an expanding
phase, as can be seen
from comparing the $\rho$ v.s. $r$
dependence to other sources.
More detailed studies of the cosmological
effects of the free axion have
been carried out \cite{lindil2,ko,gp,clw}.
Very recent work \cite{maeda} studied the special case
of a constant potential (central charge) and included the
dilaton, the axion and the spatial curvature.
These authors find
that generic solutions remain singular;
while there are some special solutions
having an infinite lifetime, they observe that these evolve
into the strong coupling region ($\phi \rightarrow \infty$)
where the theory is expected to break down, concluding that
no graceful exit can arise as a result of a free axion.
In order to complete the check of the axion's (in)ability to
produce graceful exit, it is
of interest to include an axion-dependent
potential and see if
it can yield to interpolation
between well behaved branches, evolving to and from the
weak coupling regime.

Let us therefore
consider the possibility that a (+)
branch solution can evolve
toward the region of the egg, bounce off the egg
becoming a ($-$), and
evolve away from the egg. Consider the quantity
\beq
\dot e={{6 h \dot h + 2 \dot \Lambda + \dot \rho} \over { 2 e }}
\eeq
Substituting the equations of motion,
we can rewrite this equation as
\beq
\pm \dot e=3 h^2 + {1 \over 2} \dldp =
 2 h \dot \phi - \dot h + \rho
\label{dote}
\eeq
with the sign chosen according to the branch.
Integrating this between two
times $t_0<t_1$ along a trajectory
remaining on a single branch yields
\beq
\pm(e(t_1)-e(t_0))+h(t_1)-h(t_0)=2
\int_{\phi(t_0)}^{\phi(t_1)} h d\phi
  + \int_{t_0}^{t_1} \rho dt
\label{diffint}
\eeq
This equation represents one of the trajectory equations
in implicit (integral) form.
It can be thought of as a
constraint on trajectories, and will be the key to
the proof of the
no-go theorem for (+) to ($-$) branch changing
using the egg, much like
in the axionless case \cite{lastkmo}.

Next we need to determine the direction of $\phi$ flow around
the egg. To visualize the phase space,
one can think of the $h$ axis
as a `vertical', all the other variables
lying in a `horizontal'
hyperplane. Then for a point in the phase
space to be `above'
or `below' the egg (lying on a line parallel
to the $h$ axis
and intersecting the egg),
we need $2 \Lambda + \rho \le 0$,
as we can see from (\ref{eggdef}). But solving (\ref{dotphi})
for the condition $\dot \phi=0$ shows
that $6 h^2=2 \Lambda + \rho$,
implying that $2 \Lambda + \rho \ge 0$.
Combining these two statements,
we see that trajectories cannot reverse
the direction of $\phi$ flow
above or below the egg. Thus the value of
the dilaton $\phi$
is increasing for the
trajectories above the egg (where $h>0$)
and decreasing for
the trajectories below it (where $h<0$).
This leads to the conclusion
that the contribution of the first integral
on the rhs of equation
(\ref{diffint}) is positive for such trajectories,
as it represents
the area of the projection of a trajectory on the
$h-\phi$ plane. The second
integral is also positive by the definition of $\rho$.

Finally, to complete the no-go theorem, we establish the
following two lemmas. First, we show that
no ($-$) branch trajectory leaving the top surface of the
egg can pass the boundary of the region
vertically above and below the egg (defined
by the condition that $2 \Lambda + \rho=0$). To see
this, let $t_0$ be the instant when the trajectory leaves
the egg and $t_1$ the instant
when it reaches the vertical boundary.
Clearly $e(t_0)=0$, and since the vertical boundary of the egg
region is defined by $2 \Lambda + \rho=0$, we have
$e(t_1)=\sqrt{3} h(t_1)$. Inserting this into the
lhs of the ($-$) branch version of (\ref{diffint}), we find
\beq
lhs=-\sqrt{3} h(t_1) + h(t_1) - h(t_0)
\eeq
This is clearly nonpositive (recall $h \ge 0$
since we are above the egg),
contradicting our conclusion that the
rhs is positive. An analogous argument shows that a
(+) trajectory passing the boundary and entering
the region below the egg cannot hit it.
The result is in fact just the
time-reversal of the previous statement.

{}From these two lemmas we can conclude that
egg is {\it{generically}} incapable of changing
(+) branch solutions to ($-$).
A (+) trajectory entering above the egg
must leave the region of the egg
as a (+), from the first lemma, which
prohibits the conversion
to a ($-$). Further, a (+) trajectory entering
below the egg must also leave as (+), since it cannot hit the
egg at all. We qualifiy this incapability as generic since
there is still a finely tuned set of trajectories that
may evade the proof, namely those that touch the egg
neither on the top surface nor the bottom, but at
the egg boundary in the $h=0$ hyperplane. Thus there remains
the possibility that a (+) trajectory could hit one of
these boundary points, convert to a ($-$) and then
either leave the egg region or pass underneath the
egg.

To see that this cannot happen, consider the quantity
(\ref{dote}) at a point touching
the egg in the $h=0$ hyperplane,
$\pm \dot e={1 \over 2} \dldp$. Because $e$ provides
a measure of distance to the egg ($e$ is negative
inside the egg and positive outside), we see that points
where $\dldp < 0$ repel the ($-$)
solutions and attract the (+) ones,
and conversely for $\dldp > 0$. Thus
the only case that could produce an exception to our no-go
theorem is  $\dldp \le 0$.

To consider this case, we define $f=2 \Lambda + \rho$.
As we pointed
out earlier, we have $f<0$ at points on
trajectories which are vertically
above or below the egg. At a point where
a trajectory could hit
the egg in the $h=0$ plane we must have $f=0$.
Now, we compute the time
derivative of this function
\beq
\dot f=2 \dot \Lambda + \dot \rho = 2 \dldp \dot \phi +
  2 \dlda \dot a + \dot \rho =
  2 \dldp \dot \phi - 6 h \rho
\eeq
using (\ref{ee2}). From (\ref{ee3})
we see that $\dot \phi=0$ at this
point, thus $\dot f=0$ also. Differentiating
again and making
further use of the equations of motion
shows that at this point
\beq
\ddot f = 2 \dldp \ddot \phi - 6 \dot h \rho =
 - 6 \rho^2 + 6 \rho \dldp - (\dldp)^2
\eeq
Considering $f$ as a function of time along the trajectory,
we notice that if $\ddot f<0$, then $f$ is concave and
must have been negative immediately before and after the
egg hit at $f=0$. As we have shown above that we need
only worry about the case where $\dldp \le 0$, we see that
if either $\dldp$ or $\rho$ are non-zero the trajectory
cannot escape the region directly above or
below the egg. Referring to
(\ref{doth}) we see that $\dot h>0$.
Thus these trajectories are (+)
branch trajectories coming from under the egg and
emerging above it as ($-$). However,
because we have already shown that
($-$) trajectories can not exit to the right of the egg,
we see that this case is in fact ruled out.

This argument breaks down in the special case $\dot a=0$
($\rho=0$) and $\dldp=0$. To examine the behavior here
we need to find the lowest non-vanishing
derivatives of $h$, $e$
and $f$. After some calculation we find
\beq
h^{(3)} = e^{-4 \phi} (2 (\dlda)^2
+ {1 \over 2} \dlda \dldpda)
\eeq
\beq
\pm e^{(3)} = -{1 \over 2} e^{-4 \phi} \dlda \dldpda
\eeq
\beq
f^{(6)} = -10 (\dlda)^2 (24 (\dlda)^2
+ 12 \dlda \dldpda + (\dldpda)^2)
\eeq
where $(n)$ indicates the $n$th
derivative with respect to time.
To have the (+) attracted to the
egg we need $e^{(3)} \le 0$.
This forces $h^{(3)} \ge 0$
and $f^{(6)} \le 0$ with equality
holding only in the case where $\dlda=0$.
If $\dlda=0$ then
the trajectory passing through
the point discussed above
with $h=e=f=0$ is just a constant
and so does not represent a branch change. This point is
a generalization of the exceptional
fixed points discussed
in \cite{lastkmo}. Otherwise we again
have a ($-$) branch entering the region above the egg from
which cannot escape without touching the egg again.

This completes the proof that a (+) branch, originating outside
of the egg region, cannot use it to convert to a
($-$) branch solution outside of the egg region.
As a consequence, we see that the axions cannot catalyze
branch changing and graceful exit even when endowed with
a potential.

\section{Conclusion}

We have investigated the graceful exit problem in
Pre-Big-Bang inflation with the dynamics governed
by stringy action including the axion field, as well as
a potential dependent on both the axion and the dilaton.
Although the non-interacting axion can asymptotically
link up distinct solutions of the dilaton-metric system,
in such cases the related solutions
still belong to the same branch, one past-collapsing and the
other future-expanding \cite{clw}. One
could have hoped that with the
addition of an axion potential,
actual branch changes could have
been induced, leading to the resolution of the graceful
exit problem. Specifically, one
could have expected the violations
of the positivity arguments employed in the integral formula
(\ref{diffint}), or the creation of qualitatively new pathologies
leading to at least isolated non-singular perturbative
solutions. This behavior could not be ruled out solely on the
grounds of singularity theorems, as we have shown previously.
Our analysis demonstrates that the axion in fact cannot facilitate
branch changing, and lead to
graceful exit. We thus extend our earlier
result, stating that no (+) branch
solution (generically regular
in the past) can evolve into a ($-$) branch
solution (generically
regular in the future), to hold even
in the presence of axions with
arbitrary interactions.
Thus the axion can be safely
ignored in all expanding universes,
as it does not change
their qualitative behavior.

In light of this result, we see that the only hope
to resurrect the Pre-Big-Bang
inflation is to show that higher order
$\alpha'$ corrections asymptotically link dual solutions,
and lead to nonsingular
cosmologies. This conjecture is partly supported
by the nonsingular solutions presented in \cite{art}, but
the evidence is hardly conclusive because the solutions are
non-perturbative in $\alpha'$, indicating their
sensitivity to even higher order corrections,
present in string theory. A more
comprehensive approach as advocated in \cite{kk} is in order,
where one should construct an exact conformal field theory
and study all the higher order corrections systematically.
Indeed, these authors present certain models which can be
interpreted as anisotropic universes and
which possess the desired
duality-related asymptotia. However,
the final answer should
be based on an isotropic and
homogeneous model, which is still
lacking.

\vskip 0.7truecm
\noindent {\bf Acknowledgements}
\vskip 0.7truecm
We would like to thank M. Gasperini,
R. Khuri, E. Kiritsis, and G.Veneziano for helpful
conversations. This work was supported in part by  DOE grant
DE-FG02-94ER40823, and in part by NSERC of Canada.
NK was also supported
in part by an NSERC postdoctoral fellowship.

\end{document}